\documentclass[12pt,preprint]{aastex}
\usepackage{multirow}
 \newcommand{\beq}{\begin{equation}}
 \newcommand{\eeq}{\end{equation}}
 \newcommand{\beqn}{\begin{eqnarray}}
 \newcommand{\eeqn}{\end{eqnarray}}
 
 \newcommand{\no}{\noindent}

\newcommand{\pa}{\partial}

\begin{document} 
\title{Studying Dynamical Models
of the Core Galaxy NGC 1399 with Merging Remnants
} 

\author{Li-Chin Yeh$^{1}$ and Ing-Guey Jiang$^{2}$}

\affil{
{$^{1}$Institute of Computational and Modeling Science,}\\
{National Tsing-Hua University, Hsin-Chu, Taiwan}\\ 
{$^{2}$Department of Physics and Institute of Astronomy,}\\
{National Tsing-Hua University, Hsin-Chu, Taiwan}
}

\email{jiang@phys.nthu.edu.tw}

\begin{abstract} 
An investigation on the possible dynamical models
of the core galaxy NGC 1399 is performed. 
Because early-type galaxies are likely to be formed through merging
events, remnant rings are considered in the modeling process. 
A numerical survey over three parameters
is employed to obtain the best-fit models
that are completely consistent with observations. 
It is found that the inner slope of dark matter profile 
is a cuspy one for this core galaxy.
The existence of remnant rings in best-fit models indicates 
a merging history.
The remnant ring explains the flatten surface brightness,
and thus could be the physical counterpart of the core structure of 
NGC 1399.
\end{abstract}

\newpage
\section{Introduction}

NGC 1399 is a giant elliptical galaxy located near the center of 
the Fornax cluster. According to Dullo \& Graham (2014),
it is classified as a core galaxy with a core radius 202 pc.
As cosmological N-body simulations showed that the density profiles
of dark halos can be approximated by a centrally cuspy function,
i.e. NFW profile (Navarro, Frenk, \& White 1997), it is not clear 
why some early-type galaxies such as NGC 1399 could have
a central core.

The most popular scenario to form a core is through the
gravitational sling-shot ejections of stars 
(Milosavljevic \& Merritt 2001, Merritt 2006)
by the inspiralling binary super-massive black hole (SMBH) near the center 
during merging processes of galaxies.
If this does occur, as the dark matter particles  
also experience the gravitational sling-shot ejections,
it is likely that the galactic central 
dark matter distribution will get changed. 
Further, the masses of stars and dark matter particles 
are different, the luminous and dark parts of galaxies
shall settle to different distributions in the end.
Thus, it is very important to relax the usual assumption that
``mass traces light'' and investigate the 
dark matter profiles near the centers of galaxies
through dynamical modeling. 

On the other hand, there are still problems for this sling-shot scenario
of core formation. Because each merger can only deplete 
about 0.5 SMBH mass from the center (Merritt 2006), 
a larger number of mergers are 
needed for those giant galaxies with big cores.
For NGC 1399, the data of Table 4 in Dullo \& Graham (2014)
implies that there should be about 16 mergers if the
core was formed through the sling-shot scenario.
However, according to the estimations by Conselice et al. (2002),
it is unlikely that a galaxy would experience more than 5 mergers.
In addition, investigating on the properties of mergers 
make Khochfar \& Burkert (2005) and Naab et al. (2006) 
claim that mergers could lead to slowly rotating boxy remnants.
A boxy remnant could become the main part of a newly formed 
giant elliptical galaxy. The slow rotation implies that 
some part of the remnant owns 
certain amount of angular momentum and it is possible that 
this part might become a disky remnant ring
hidden in the galaxy. This remnant ring could contribute 
to the surface brightness of a core structure.  
 
Therefore, it motivates us to study whether 
a model containing a stellar remnant ring could fit   
the observational constraints well.
That is, in this paper, we will investigate the existence of dynamical 
models which include both the main stellar part and the remnant stellar part.
The main stellar part is assumed to be spherically distributed and the remnant 
could be a ring-like structure which is located on the orbital
plane of spiralling binary SMBH during the merging of galaxies.
 
As for previous work,
Saglia et al. (2000) presented a detailed study on the mass distributions
of NGC 1399. Since then, previous studies about NGC 1399 
focused on either the dynamical structure 
of the outskirts or the mass measurement of central SMBH. 
For example, employing 
the planetary nebula data, Napolitano et al. (2002)
studied the velocity structures of outer regions of NGC 1399 and 
concluded that the interactions from nearby galaxies are important.
In addition, Schuberth et al. (2010) and Samurovic (2016)
also addressed the dynamics of outer parts through the kinematic data 
of globular clusters.    
Schulz et al. (2016) even estimated the star formation rates
of NGC 1399 from the ages of globular clusters. 
Samurovic \& Danziger (2006) established the total dynamical mass
of NGC 1399 through both studies in X-rays and kinematics 
of globular clusters.
Moreover, focusing on the very central region of NGC 1399, 
Houghton et al. (2006) and Gebhardt et al. (2007) 
provided precise estimations of the SMBH mass.  

In addition to employing Jeans equations as in Samurovic (2016), 
to construct a model of early-type galaxies, 
the orbit-based method (Schwarzschild 1979, 1993) is often used.
The superposition of orbits with different weights 
in a fixed galactic potential is employed to 
build theoretical models which could fit the 
observed brightness profiles and kinematics data.
This method has been widely used in many papers such as Rix et al. (1997)
and van der Marel et al.(1998).
On the other hand, Syer \& Tremaine (1996)
proposed a particle-based method in which the weights 
of particles could be changed when these particles 
were proceeding along their fixed orbits,
see the discussions
about these methods in McMillan \& Binney (2012).

Because the main goal of our work here is to investigate 
the dark matter density profile
and study the existence of a stellar remnant ring 
of the core galaxy NGC 1399,
in order to explore possible profiles of 
both luminous and dark matter within the framework 
of standard dynamical systems and have the freedom
to consider the possible stellar remnant ring, 
different from Schwarzschild (1979)
and Syer \& Tremaine (1996),
the usual particle-based dynamical simulations will be used here. 
In fact, N-body simulations were used to construct models
of globular clusters (Baumgardt 2017).
However, for galaxies, many more particles will be needed
and it would be very difficult to tune model parameters
to fit observational data if expensive N-body simulations 
are employed.
The model in Kandrup et al. (2003) gives an 
excellent example that using fixed galactic total potential,
N-body test-particle simulations can lead to 
stellar structures that fit with observations well.
  
In this paper, we set the total galactic potential to be fixed, and 
use N-body particles to represent the stellar part.
With reasonable choices of initial conditions of stellar particles,
the theoretical surface brightness and velocity dispersion can be obtained 
when these particles approach to a quasi-equilibrium.
The model results will be compared with the observational data.
In order to consider the possible merging remnants,
further stellar particles
will be added in models in a way to improve the fitting with 
the observational surface brightness and velocity dispersion.
   
The principle procedure of seeking an equilibrium through an N-body 
dynamical system is
described in Section 2.
The model details are in Section 3 and 
the results are in Section 4.
Concluding remarks are in Section 5. 
 
\section{The Procedure}

N-body dynamical systems are often employed to model galaxies
(Wu \& Jiang 2009, 2012, 2015). 
For an N-body system, any particle with coordinate $(x, y, z)$ 
follows the equations of motion given below:
\beq\left\{
\begin{array}{ll}
\frac{d^2 x(t)}{d t^2} =  -  \frac{\pa \Phi(x,y,z,t)}{\pa x},\\
\frac{d^2 y(t)}{d t^2} =  -  \frac{\pa \Phi(x,y,z,t)}{\pa y},\\
\frac{d^2 z(t)}{d t^2} =  -  \frac{\pa \Phi(x,y,z,t)}{\pa z},\\
 \end{array}  \right. \label{eq:eq1}
\eeq
where $\Phi(x,y,z,t)$ is the total potential of this system.
In a realistic N-body simulation, 
$\Phi(x,y,z,t)$ is evaluated repeatedly at each time step.
However, in this paper, in order to be more efficient in searching 
for equilibrium models constrained by observational data,
the total galactic potential is fixed
and the SMBH is represented by a point mass at the center.
The total potential is set to be 
a summation of galactic total potential and SMBH potential.

Because the image of NGC 1399 is circularly symmetric,
for convenience, the total galactic potential 
is assumed to be spherical as in 
Samurovic (2016). 
All particles, which represent the stellar part 
are governed under the parameterized total potential. 

For particles' initial positions,
due to the fact that NGC 1399 is a core galaxy (Dullo \& Graham 2014),
these stellar particles 
are set to follow a spherical core-double-power-law density distribution.
The initial velocities are set based on a
reasonable function of anisotropic parameters.

After initial positions and velocities are given, 
all particles' orbits will be determined through the above
equations of motion. 
When the system settles into a quasi-steady distribution
at a particular time, i.e. $t_{end}$, 
all particles' positions and velocities are recorded 
at this snapshot.
The corresponding stellar surface brightness and velocity dispersion
are calculated and compared with the observations.

To investigate the existence of a stellar remnant ring,
an axisymmetric stellar part is then considered. 
The initial positions and velocities of axisymmetric
particles are chosen in such a way that 
after they are integrated to $t_{end}$ and added into the system,
the overall $t_{end}$ snapshot can have a better fitting with the
observations. 
That is, through particle superposition,
an axisymmetric stellar part 
is added in a way to improve the surface-brightness 
and kinematic fittings.  
It is called the remnant ring hereafter, which could be a  
structure left over during the formation of this system.

\section{The Models}

Following the procedure mentioned in the previous section, 
further details are described here.
 
\subsection{The Observational Constraints}

Because the observed surface brightness profile
was well fitted by a core-Sersic law in 
Dullo \& Graham (2014), that core-Sersic law 
is used as our observational constraint for
the surface brightness of NGC 1399.
The long-slit velocity dispersion data  
from Graham et al. (1998) covers the whole core region of NGC 1399 
with reasonable resolutions. Thus, it is employed as
our observational constraint as it fits our goal to determine 
the luminous and dark matter distributions around this region.
 
In addition, the mass of central SMBH was determined 
to be $5.1\pm 0.7 \times 10^{8} M_{\odot}$ 
in Gebhardt et al. (2007) 
and $1.2^{+0.5}_{-0.6} \times 10^{9} M_{\odot}$
in Houghton et al. (2006). 
The above two values are actually consistent with each other.
Due to the fact that very central kinematics is measured in 
Houghton et al. (2006), we set the mass of SMBH to be 
$1.2\times 10^{9} M_{\odot}$ in this paper. 
 
\subsection{The Units}

The mass unit is $1.2 \times 10^{12} M_{\odot}$,
the length unit is kpc, and the time unit is
$4.3 \times 10^5$ years. With these units,
the gravitational constant $G=1$.  

\subsection{The Density Profiles}

The central galactic total density profile, i.e. 
the stellar part plus dark matter, 
follows the double-power law as 
\beq
\rho_{\rm total}(r)=\rho_{t}\left(\frac{r}{r_{tb}}\right)^{-\gamma}
\left\{1+\left(\frac{r}{r_{tb}}\right)^{\alpha}\right\}
^{\frac{\gamma-\beta}{\alpha}},\label{rho_t}
\eeq
\no where $\rho_{t}$ is a constant and the scale length 
$r_{tb}$ is the total break radius. 
When $r \gg r_{tb}$, the above becomes a power law with index
$-\beta$. For $r \ll r_{tb}$, 
$\gamma$ is the inner cusp slope 
of this density profile. 

In order to include different possible slopes of the central cusp, 
we choose six profiles with different values of $\alpha$, $\beta$, 
and $\gamma$ as listed in Table 1. 
In Table 1,  
Profile A and B are for $\gamma=0$; Profile C and D are for $\gamma=1$;  
Profile E and F are for $\gamma=2$.
These profiles are equivalent to some of those with analytic potentials
discussed in Zhao (1996). 

\newpage 
{\centerline {\bf Table 1. The density profiles}}
    \begin{center}
       \begin{tabular}{|c|c|c|c|c|}\hline
Profile & $\alpha$ & $\beta$ &  $\gamma$  &  $\rho_{\rm total}(r)$ \\ 
\cline{1-5}
 A  & 2 & 3 &0 & $\rho_t\left\{1+\left(\frac{r}{r_{tb}}\right)^2\right\}
^{-1.5}$    \\ \cline{1-5}
 B  & 2 & 5 &0 & $\rho_t\left\{1+\left(\frac{r}{r_{tb}}\right)^2\right\}
 ^{-2.5}$    \\ \cline{1-5}
 C  & 1 & 4 &1 & $\rho_t\left(\frac{r}{r_{tb}}\right)^{-1}
\left\{1+\frac{r}{r_{tb}}\right\}^{-3}$   \\ \cline{1-5}
 D  & 2 & 5 &1 & $\rho_t\left(\frac{r}{r_{tb}}\right)^{-1}
\left\{1+\left(\frac{r}{r_{tb}}\right)^2\right\}^{-2}$    \\ \cline{1-5}
 E  & 1 & 4 &2 & $\rho_t\left(\frac{r}{r_{tb}}\right)^{-2}
\left\{1+\frac{r}{r_{tb}}\right\}^{-2}$     \\ \cline{1-5}
 F  & 1 & 6 &2 & $\rho_t\left(\frac{r}{r_{tb}}\right)^{-2}
\left\{1+\frac{r}{r_{tb}}\right\}^{-4}$     \\ \cline{1-5}
\end{tabular}
    \end{center}

In addition to the above, due to the fact that 
the velocity dispersion remains to be nearly a constant
from $R=$ 2 kpc to $R=$ 6 kpc, 
more dark matter shall be present around that region 
(Saglia et al. 2000).  In order to have more 
galactic mass out of $R=$ 2 kpc, another power-law 
profile is employed. Thus, 
\beq
\rho(r)=\left\{\begin{array}{ccc}
&\rho_{\rm total}(r)  & \quad{\rm if} \quad  r \le r_d, \\
& cr^{p}  & \quad {\rm if} \quad r>r_d, \\
\end{array}\right.
\eeq
\no where $p$ is the power-law index,  
the constant $c=\frac{\rho_{\rm total}(r_d)}{(r_d)^p}$,
and $r_d$ is the transition radius.

In order to search for the best model, 
several grids of parameters are considered.
There would be six central galactic total density profiles, 
10 values of $r_{tb}$, 10 values of $p$,
and four values of $r_d$.
That is, under Profile A, B, C, D, E, F,
we have $r_{tb}$ = 0.3 to 1.2 with step 0.1,
$p$ = -2.1 to -1.2 with step 0.1, and
$r_d$ = 2.5, 3.0, 3.5, 4.0.
Thus, totally 2400 models would be calculated in this paper.
 
Note that one SMBH is placed at the coordinate center.
In addition to the total galactic potential 
derived from the above density profiles,
one SMBH's gravitational potential is added.

Since NGC 1399 is a core galaxy,  
$10^6$ stellar particles are set to follow an initial 
distribution as Profile A.
Thus,
\beq
\rho_{\rm star}(r)=
\rho_s 
\left\{1+\left(\frac{r}{r_{sb}}\right)^2\right\}^{-1.5},
\label{rho_star}
\eeq
where $\rho_s$ is a constant and 
the scale length $r_{sb}$ is the stellar break radius.

It is found that we need to set $r_{sb}=1.0$, in order to 
have enough stellar particles at the main part of the system.
At this radius, a typical velocity is estimated to be $v_{sb}=0.1$,
and a dynamical time $t_d \equiv r_{sb}/v_{sb}$=10.
Employing the leapfrog integrator (Binney \& Tremaine 2008),
the orbits of all particles are calculated from 
$t=0$ to $t= t_{end} \equiv 10 t_d$.

\subsection{The Initial Velocities}

The assignment of particles' initial velocities is described here.
The velocity magnitude is set to be the circular velocity with respect to 
the galactic total mass (excluding SMBH mass) at the particle's position,
i.e. $v_{cir}(r)$. This velocity is divided into two components,  
the radial velocity $v_r$ and the tangential velocity $v_t$. 
Thus, $v_{cir}^{2}(r) = v_t^2 + v_r^2$.
 
Because the anisotropy parameter $\beta_a$ (Binney \& Tremaine 2008) 
satisfies
\beq
\beta_a = 1 - \frac{\bar v_t^2}{2 \bar v_r^2 },
\eeq
where $\bar v_t^2$ is the average tangential velocity over particles,
and  $\bar v_r^2$ is the average radial velocity over particles.
Theoretical models of galaxy formation show that $\beta_a$ 
generally increases with radius (Binney \& Tremaine 2008).
In order to choose our initial velocities properly to make that 
$\beta_a$ does increase with radius approximately, 
we create a function 
$\beta_p(r)$ which has a value -1 at $r= 0$ kpc 
and becomes 1 at $r=6$ kpc (Note we set the outer boundary to be
6 kpc for the system considered in this paper). 
The simplest nonlinear polynomial function which has this
mathematical property is  
\beq
\beta_p(r)= - \frac{1}{18}r^2 + \frac{2}{3}r -1.
\eeq
 
Because the magnitudes of radial and tangential velocities
satisfy
\beq
v_t^2+v_r^2=v_{cir}^2\equiv \frac{GM(r)}{r},
\eeq
where $M(r)=4\pi\int_0^r\bar{r}^2\rho(\bar{r})d\bar{r}$, 
and we set $v_t^2=2(1-\beta_p(r))v_r^2$,
they can be obtained as
\beq
\left\{\begin{array}{ll}
& v_r= \sqrt{\frac{GM(r)}{ [ (2-\frac{1}{3}r)^2 +1 ] r} }\\
& v_t= |(2-\frac{1}{3}r)|v_r.
\end{array}\right.
\eeq

As for the directions, the radial velocity is set to
be either leaving from the system center (outward) 
or going to the system center (inward) with equal probability.
The direction of tangential velocity is set to be isotropic.
The particle velocity is thus determined after 
the radial and tangential velocities are settled.
 
As an example, Fig. 1 gives the information of initial velocities
of a model with parameters $r_{tb}=1.0$, $p=-1.3$, and $r_d=3.0$.
The anisotropy parameter $\beta_a$
as a function of radius for Profile A-F  are presented 
in Fig. 1(a).
The average $v_t^2$ (dashed curves) and the 
average $v_r^2$ (solid curves) as functions of radius
for Profile A-F are shown in Fig. 1(b).

\subsection{The Surface Brightness and Velocity Dispersion}

In our coordinate system, the line of sight is along the
$z$-axis, and the plane of the sky is set to be 
the $x-y$ plane. 
In order to calculate the theoretical surface brightness,
radial bins on the $x-y$ plane are set as below.
The circle with radius $R_0$ centering on the coordinate center
is the 0-th radial bin. The annulus between
$R_0$ and $R_1$ is the 1st radial bin.
In general, the $i$-th radial bin is the annulus between
$R_{i-1}$ and $R_{i}$, where $i=1,2,3,...$.
Here we choose $R_0$=0.1, and $\delta R \equiv R_{i}-R_{i-1}=0.2$ 
for any positive integer $i$.

To calculate the velocity dispersion, cells 
are further set at the long-slit positions along the $x$-axis.
The cell height is set to be $h$. Here we set $h= 0.1$. 
Within the 1st radial bin, 
the cell ranging from $\theta= -\tan^{-1}(h/R_1)$
to $\theta = \tan^{-1}(h/R_1)$
is called the right cell,
the cell ranging from $\theta= \pi -\tan^{-1}(h/R_1)$
to $\pi + \theta = \tan^{-1}(h/R_1)$
is called the left cell.

Similarly, within the $i$-th radial bin for any integer $i > 0$,
the cell ranging
from $\theta=-\tan^{-1}(h/R_{i})$  to $\theta = \tan^{-1}(h/R_{i})$
is the right cell,
the cell ranging
from $\theta=\pi-\tan^{-1}(h/R_{i})$  to $\theta = \pi + \tan^{-1}(h/R_{i})$
is the left cell. 
For NGC 1399, the observational data of velocity dispersion
along $+x$-axis and 
$-x$-axis are both available. The average values of two sides
are used as the observational values. The particles in both 
right and left cells are included in the
calculations of theoretical velocity dispersion.
 
Fig. 2 shows the surface brightness as a function of radius $R$.
The solid curve is the observational profile derived 
from the $R$-band data (Dullo \& Graham 2014).
That is
\beq
\mu(R) = -2.5\log_{10}(I(R)/I(R_b)) + \mu_{b},
\eeq
where $\mu_{b}=16.36$ is the mag at the core radius 
$R_b=0.202$ kpc, and the analytic core-Sersic law 
(Dullo \& Graham 2014)
\beq 
I(R)=I_0\left[1+\left(\frac{R_b}{R}\right)^{\alpha_n}\right]^
{{\gamma_n}/{\alpha_n}}
\exp\left[-b_n\left(\frac{R^{\alpha_n}+R_b^{\alpha_n}}
{R_e^{\alpha_n}}\right)^{1/(\alpha_n n)}\right],
\eeq
where $R_e=3.22$ kpc, $\alpha_n=2$, $\gamma_n=0.11$,
$b_n=2n-1/3$, and $n=5.6$.
Note that $I_0$ is a normalization constant which can be determined by 
$\mu_{b}=16.36$. 

With the same normalization, 
the values of theoretical surface brightness 
in our models are then determined.
In Fig. 2, the points are the average surface brightness of 
2400 models at the $t_{end}$ snapshot. 
The error bars are the corresponding standard deviations.  
Thus, for $R \ge 1$ kpc, the theoretical surface brightness of all models
generally follow the observations.
However, for $R<1$ kpc, 
the theoretical surface brightness of all models are smaller than
the observations.   

\subsection{The Remnant Ring}

An additional component presenting a remnant ring is
considered here. These added particles are called 
remnant-ring particles.
For convenience, the particles which belong to the original spherical part 
would be called the main-part particles.

The numbers of remnant-ring particles in radial bins
are decided by comparing the theoretical surface brightness
contributed from main-part particles in these radial bins 
with the observational surface brightness. 
Fig. 3(a)-(c) shows the histograms of the number of remnant-ring particles
of all models in Profile C, D, and E.
In each radial bin, the initial positions of remnant-ring particles
are uniformly distributed between two boundaries of the bin 
and within a $z$ interval
$[-z_t, z_t]$, where $z_t=R_i/60$.
This leads to a linearly increasing thickness  for the remnant ring.
The particles' radial distances on $x-y$ plane are
$R={\sqrt {x^2 + y^2}}$.
Their $x$ and $y$ components of velocities are assigned 
in a way that their orbits are circular on $x-y$ plane.   

The $z$ components of 
remnant-ring particles' velocities, 
which are along the line-of-sight,
will affect the values of theoretical velocity dispersion.
A target line-of-sight velocity distribution
in each radial bin
is set as a Gaussian distribution function with a $\sigma$
being the same as the observational velocity dispersion.
The remnant-ring particle's $v_z$ is set in a way to
make the total number of main-part and remnant-ring particles over
$v_z$ velocity bins  equal to this target distribution's integral 
in $v_z$ velocity space for each radial bin.
 
With the above initial conditions, the orbits of remnant-ring particles
are calculated from $t=0$ to $t=t_{end}$ 
through the leapfrog integrator (Binney \& Tremaine 2008).
Considering $t_{end}$ snapshots, these remnant-ring particles
would be included as part of stellar particles when
the final theoretical velocity dispersion and surface brightness
are calculated.
Thus, after remnant rings are included,
our models become complete.
As we can see from Fig.3, the numbers of remnant-ring particles  
in our models range from 50000 to 150000, 
which is about 5 to 15 percents of the main-part. 

\section{The Results}

The $t_{end}$ snapshots of all these 2400 models 
are considered,
and used to compare with observations.  
The surface brightness of these models can fit the 
observationally determined core-Sersic law very well.
There are some deviations around the outer part of the considered region 
due to very small number of particles around there.
We ignore that deviation as our study focuses on the central
profiles around the core structure.
 
We then calculate the theoretical velocity dispersion of 
these 2400 models in order to compare with the observational 
velocity dispersion.
The chi-square fitting is used to do the above comparison and 
obtain the best-fit models. We set 
\beq
\chi^2=\sum_{i=0}^{28}\frac{(O_i-T_i)^2}{O_i},
\eeq
where $O_i$ is the observational velocity dispersion
of the $i$-th radial bin 
and $T_i$ is the theoretical velocity dispersion
of the $i$-th radial bin. 
There are 29 bins, so the degree of freedom is 28
and the reduced chi-square $\chi_{\rm red}^2$= $\chi^2/28$. 
We obtain $\chi_{\rm red}^2$ for all 2400 models.

Figs. 4-6 show the values of reduced chi-square $\chi_{\rm red}^2$ 
of 400 models in Profile C, D, and E, respectively.
In general, when the $\chi_{\rm red}^2$ value is much larger than one,
the model is unlikely to be the correct one;
 when the $\chi_{\rm red}^2$ value is around one,
the model has a high probability to be the correct one.
For example, as shown in Figs. 4-6, those models with $p=-1.7$ 
are all unlikely ones. However, 
there are also plenty of possible models. 
The one with the smallest $\chi_{\rm red}^2$ is regarded
as the most likely model.
Because many models have very similar values of $\chi_{\rm red}^2$,
we decide to pick up three models, whose $\chi_{\rm red}^2$
are smaller than all the rest, as best-fit models. 

\newpage
 {\centerline {\bf Table 2.  The best-fit models} 
    \begin{center}
       \begin{tabular}{|c|c|c|c|}\hline
Model   & 1 & 2 & 3 \\ \cline{1-4}
Profile & C & D &  E  \\ \cline{1-4}
$r_{tb}$  & 0.3 & 0.6 & 0.9 \\ \cline{1-4}
$r_d$  & 3.0 & 3.0 &3.0 \\ \cline{1-4}
$p$ & -1.3 & -1.3 & -1.3\\ \cline{1-4}
remnant-ring   & 118376  & 115940 & 99347 \\
particle numbers  &  &  &  \\\cline{1-4} 
$\chi_{\rm red}^2$ & 0.47 & 0.48 & 0.38 \\\cline{1-4} 
\end{tabular}
    \end{center}

In Table 2, the parameters of these three best-fit
models are presented.
Although their central galactic total density profile 
are different, they are all centrally cuspy models.
That is, Model 1 and Model 2 have the same inner cusp slope
$\gamma=1$, while Model 3 has a larger inner cusp slope
with $\gamma=2$.
In addition, they have different total break radius, but 
their transition radius and power-law index are exactly the same.
The numbers of remnant-ring particles are about 10 percent of
the main-part stellar particles.
 
From the values of $\chi_{\rm red}^2$, it is clear that 
these three models are completely consistent with the observational data.
Fig. 7 shows the velocity dispersion as a function of $R$.  
The line with circles are for Model 1, 
the line with triangles are for Model 2, 
and the line with crosses are for Model 3.
The squares with error bars are for 
the observational data.
Fig. 8 presents the surface brightness of best-fit models. 
The circles are for Model 1,
triangles are for Model 2, 
crosses are for Model 3,
and solid curve is the observational surface brightness
set by Eqs. (9)-(10) with the same parameters. 

Moreover, the dark matter density profiles can be obtained by 
subtracting the stellar density from the galactic total density.  
Considering the spherically distributed parts, 
Fig. 9(a)
shows the dark matter density profiles (triangles),
the stellar density profiles (circles), and 
the galactic total density profiles (solid curves)
of Model 1. 
Similarly, Fig. 9(c) and Fig. 9(e) are for Model 2 and Model 3.

In addition,
the total cumulative mass as a function of radius, $M(r)$ is shown
in the right panels of Fig. 9. 
In Fig. 9(b), the dark matter (triangles), the stellar part (circles),
and the galactic total cumulative mass (solid curves) 
of Model 1 are presented.
Fig. 9(d) and Fig. 9(f) are for Model 2 and Model 3, respectively.

As can be seen in Fig. 9 that the inner cusp slope of galactic total
density profile is mainly contributed by the dark matter.
It is consistent with the results of cosmological simulations
that the dark matter profile is cuspy as the NFW profile has 
a central cusp with a power-law index $\gamma=1$.  
Furthermore, it is clear that
the results of Models 1-3 presented in Fig. 9 are very similar. 
That is, the best-fit models converge into one physical model   
after a numerical survey over several parameters and 
examining 2400 models. Therefore, this physical model 
we found could well represent the core galaxy NGC 1399.

\section{Concluding Remarks}

The structures of early-type galaxies continue to be an important subject
as they are resulted from the formation and interactions of galaxies.
The dynamical histories of galaxies could be revealed through 
the modeling of these galaxies.
It is interesting that some early-type galaxies have 
core structures given that cosmological simulations
show cuspy dark matter profiles. 
In order to address this, both the distributions of luminous matter 
and dark matter shall be determined separately through dynamical 
modeling.
 
Motivated by the core structure of central surface brightness
of early-type galaxies, through particle superposition, 
we construct dynamical models of the core galaxy NGC 1399.
The distributions of luminous and dark matter are studied
for this galaxy.  
Because early-type galaxies are likely to be products of mergers,
remnant rings are considered in our models.

A numerical survey of parameters is performed and three best-fit models
are obtained. We conclude that the inner slope of dark matter profile 
is a cuspy one for this core galaxy NGC 1399. In addition, the 
remnant ring contributes to the central surface brightness 
of NGC 1399. Thus, this implies
that the remnant ring could be a candidate 
of the physical counterpart of the
core structure of early-type galaxies.  

\section*{Acknowledgment}
We are grateful to the referee, Srdjan Samurovic, for many
important suggestions.
This work is supported in part 
by the Ministry of Science and Technology, Taiwan, under
Li-Chin Yeh's 
Grants MOST 106-2115-M-007-014
and Ing-Guey Jiang's
Grants MOST 106-2112-M-007-006-MY3.

\clearpage

\begin{figure}[hbt]
\centering
\includegraphics[width=1.1\textwidth]{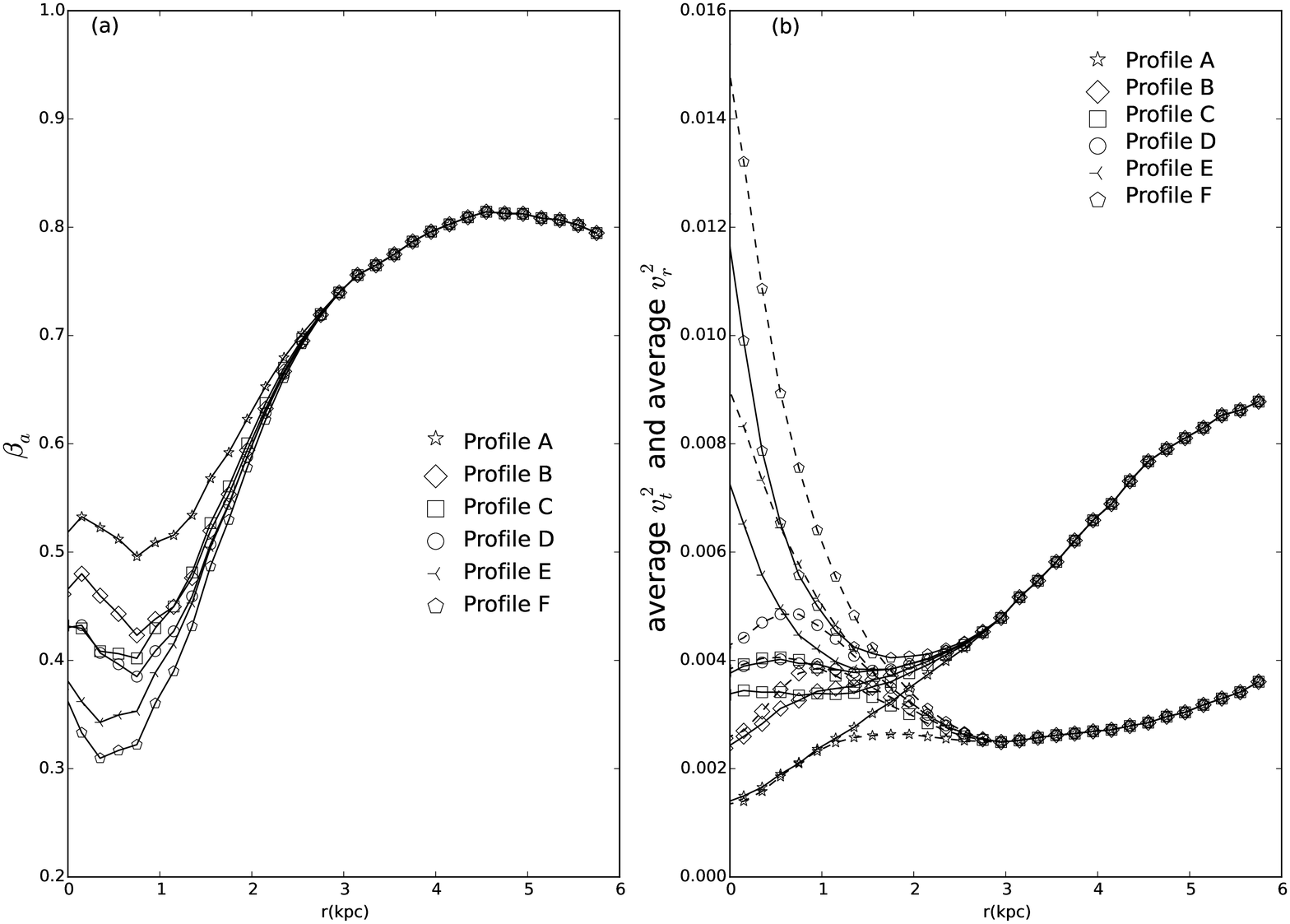}
\caption{The initial velocities. 
(a) The value of $\beta_a$ as a function of radius $r$
for Profiles A-F
(b) The average $v_t^2$ (dashed curves)
and the average $v_r^2$ (solid curves)   
as functions of radius $r$ 
for Profiles A-F.}
\label{fig:fig_1}
\end{figure}

\begin{figure}[hbt]
\centering
\includegraphics[width=1\textwidth]{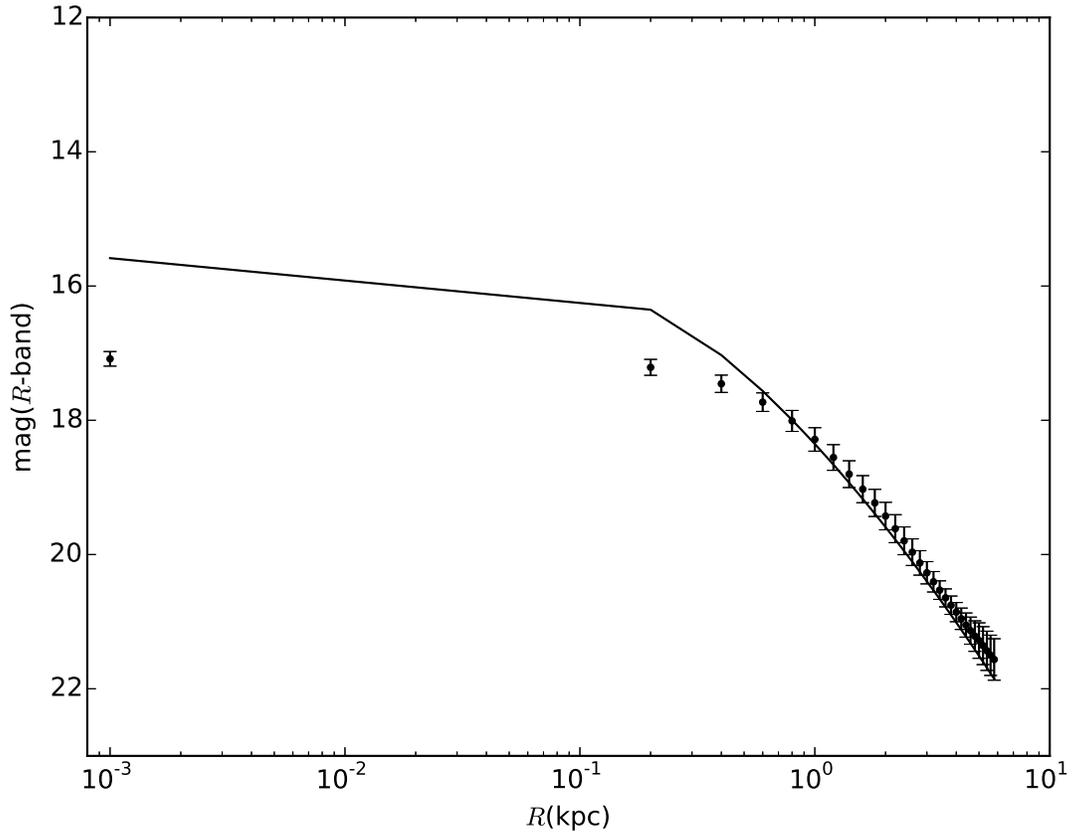}
\caption{The surface brightness as a function of radius $R$
of main-part stellar particles at the $t_{end}$ snapshot.
The average theoretical surface brightness of all 2400 models
and their standard deviations are shown by 
the points with error bars.
The solid curve is for the observation, set
by Eqs. (9)-(10) with the same parameters as written in the main text.
}
 \label{fig:fig_2}
\end{figure}

\begin{figure}[hbt]
\centering
\includegraphics[width=1.0\textwidth]{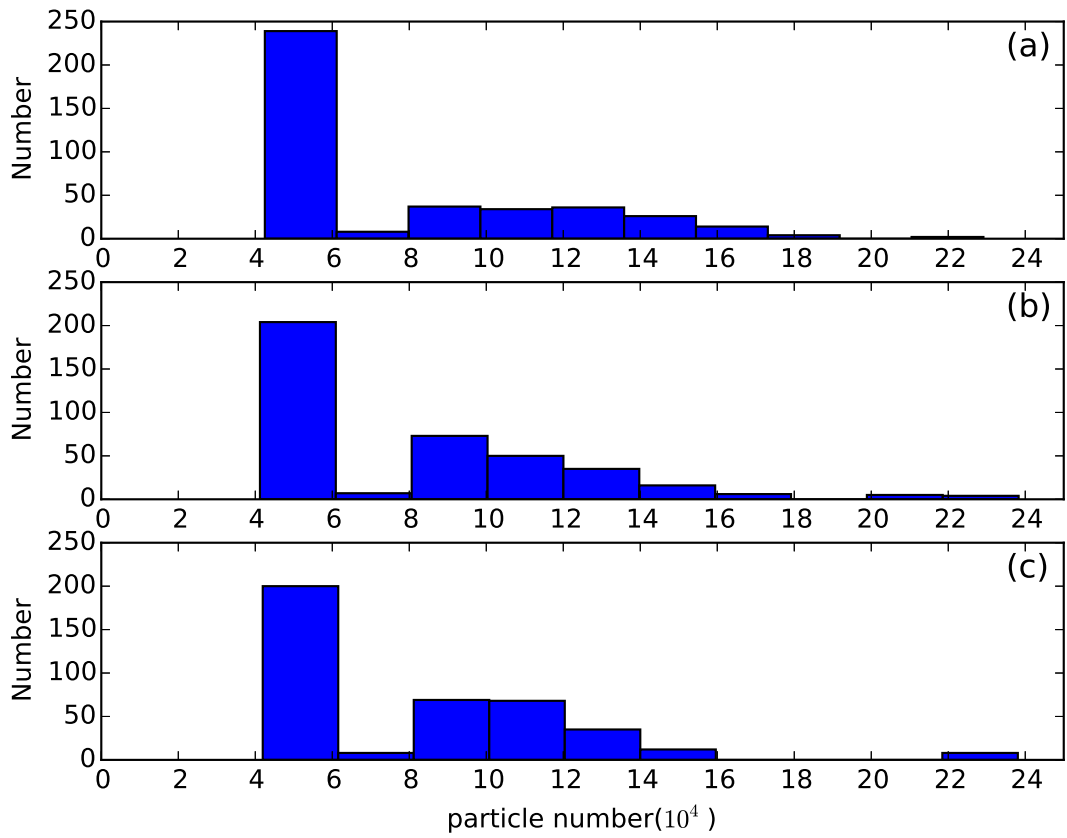}
\caption{The histograms of particle numbers of the remnant ring  
in 400 models of three profiles. Panel (a) is for Profile C,
Panel (b) is for Profile D, and Panel (c) is for Profile E.}
 \label{fig:fig_3}
\end{figure}

\begin{figure}[hbt]
\centering
\includegraphics[width=1\textwidth]{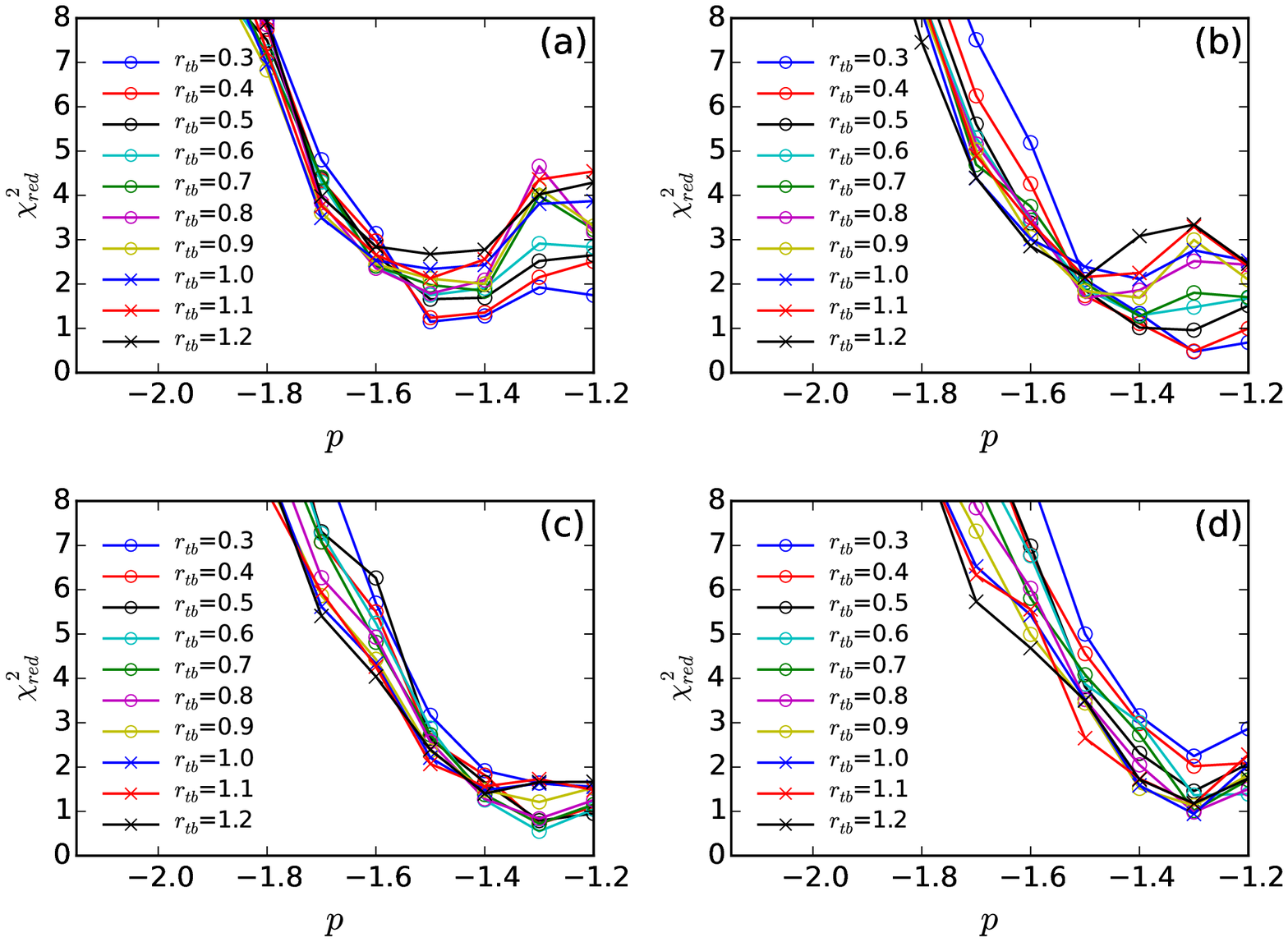}
\caption{The value of reduced chi-square $\chi_{\rm red}^2$ 
as a function of $p$ 
for different values of $r_{tb}$ in Profile C. 
Panel (a) is for $r_{d}=2.5$, Panel (b) is for $r_{d}=3.0$, 
Panel (c) is for $r_{d}=3.5$, and Panel (d) is for $r_{d}=4.0$.
}
\label{fig:fig_4}
\end{figure}

\begin{figure}[hbt]
\centering
\includegraphics[width=1\textwidth]{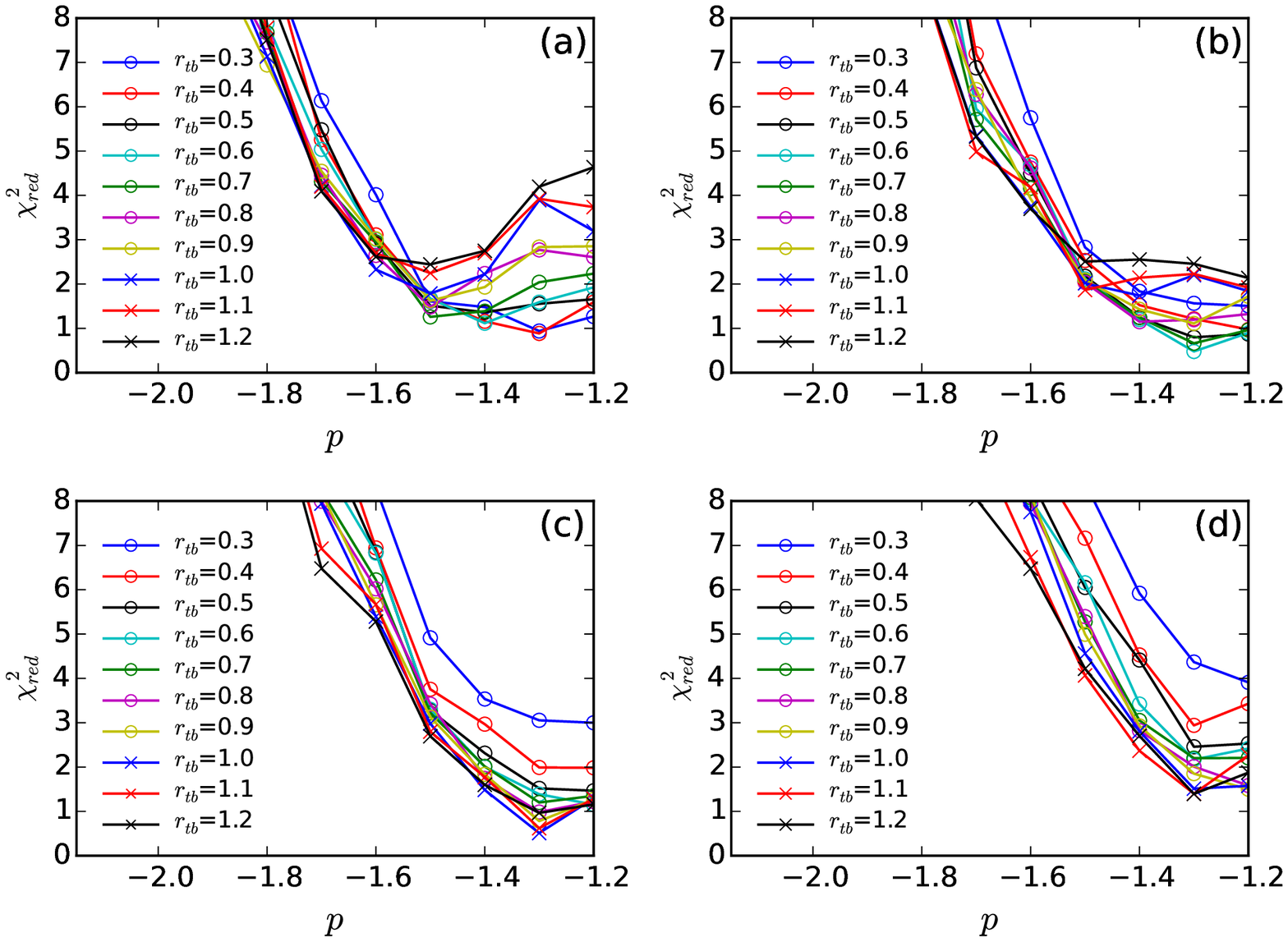}
\caption{The value of reduced chi-square  $\chi_{\rm red}^2$ 
as a function of $p$ 
for different values of $r_{tb}$ in Profile D. 
Panel (a) is for $r_{d}=2.5$, Panel (b) is for $r_{d}=3.0$, 
Panel (c) is for $r_{d}=3.5$, and Panel (d) is for $r_{d}=4.0$.
}
 \label{fig:fig_5}
\end{figure}

\begin{figure}[hbt]
\centering
\includegraphics[width=1\textwidth]{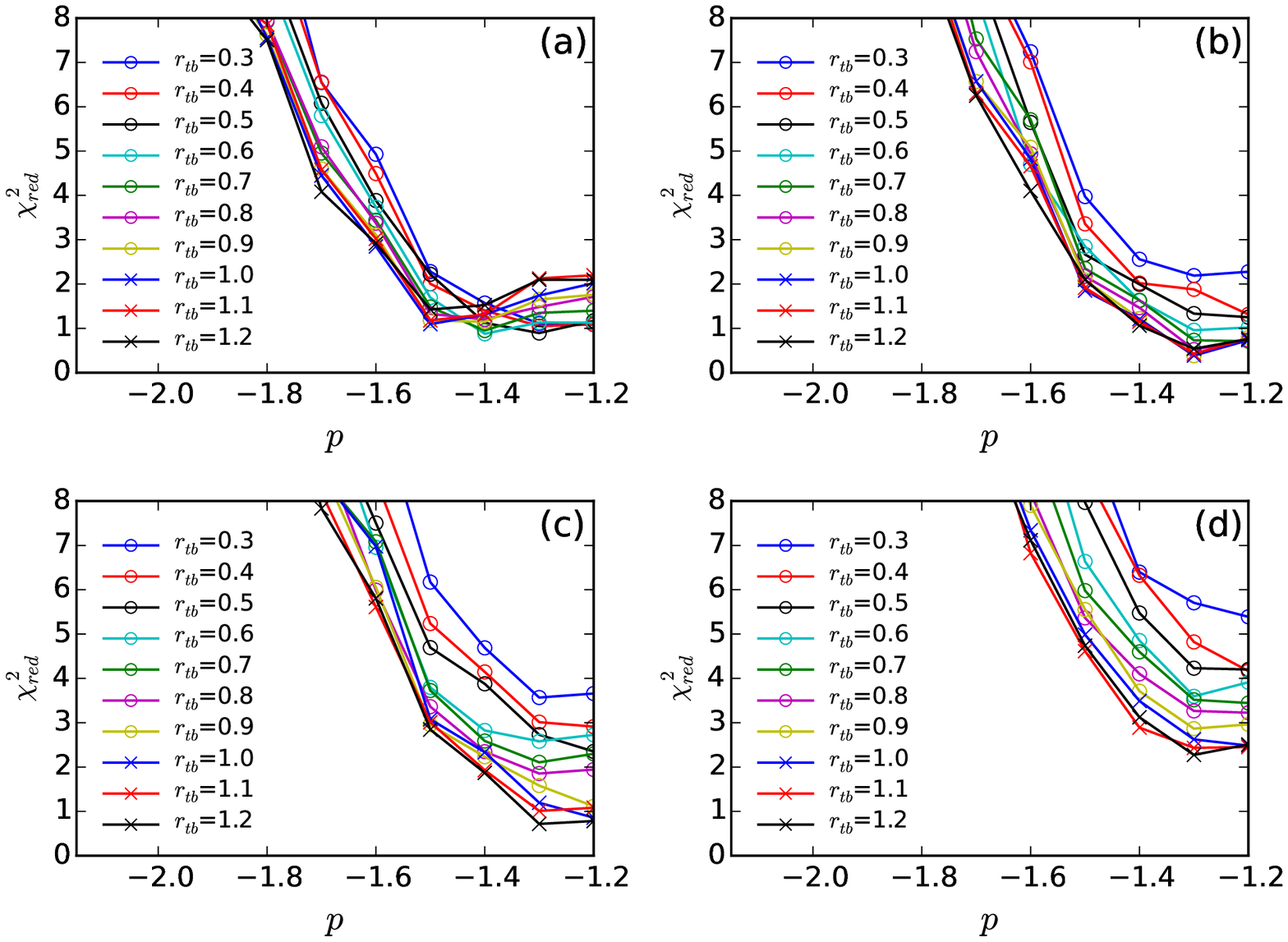}
\caption{The value of reduced chi-square $\chi_{\rm red}^2$ 
as a function of $p$ 
for different values of $r_{tb}$ in Profile E. 
Panel (a) is for $r_{d}=2.5$, Panel (b) is for $r_{d}=3.0$, 
Panel (c) is for $r_{d}=3.5$, and Panel (d) is for $r_{d}=4.0$.
}
\label{fig:fig_6}
\end{figure}

\begin{figure}[hbt]
\centering
\includegraphics[width=1\textwidth]{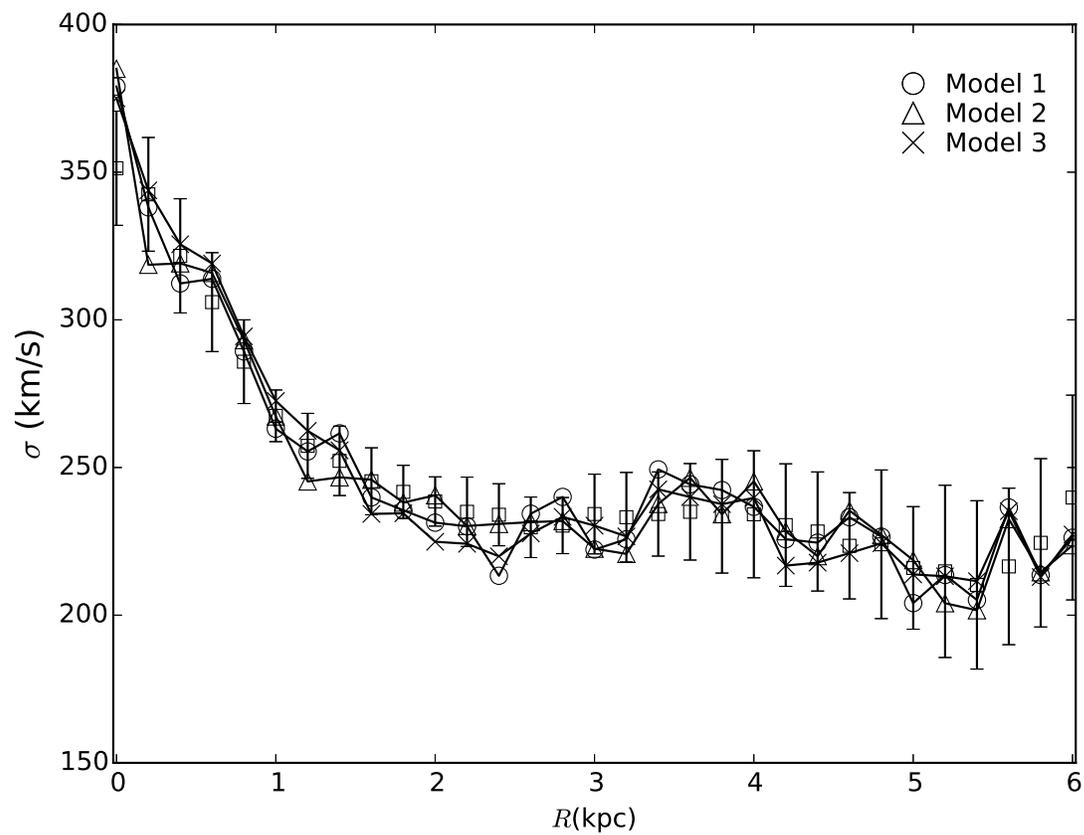}
\caption{The velocity dispersion as a function of $R$ 
for three best-fit models. 
The line with circles are for Model 1, 
the line with triangles are for Model 2, 
and the line with crosses are for Model 3.
The squares with error bars are for 
the observational data.}
 \label{fig:fig_7}
\end{figure}

\begin{figure}[hbt]
\centering
\includegraphics[width=1\textwidth]{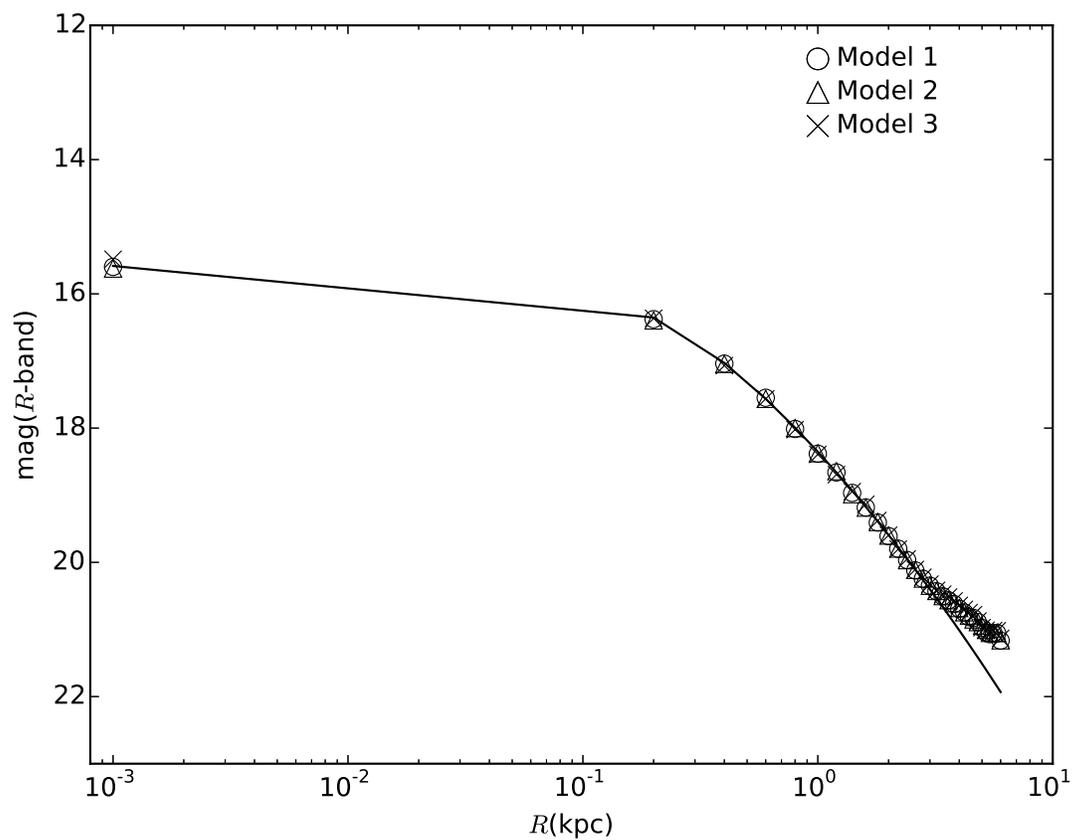}
\caption{The surface brightness as a function of $R$  
for three best-fit models. 
The circles are for Model 1, 
the triangles are for Model 2, 
and the crosses are for Model 3.
The solid curve is for the observation, set
by Eqs. (9)-(10) with the same parameters as written
in the main text.}
 \label{fig:fig_8}
\end{figure}

\begin{figure}[hbt]
\centering
\includegraphics[width=1.1\textwidth]{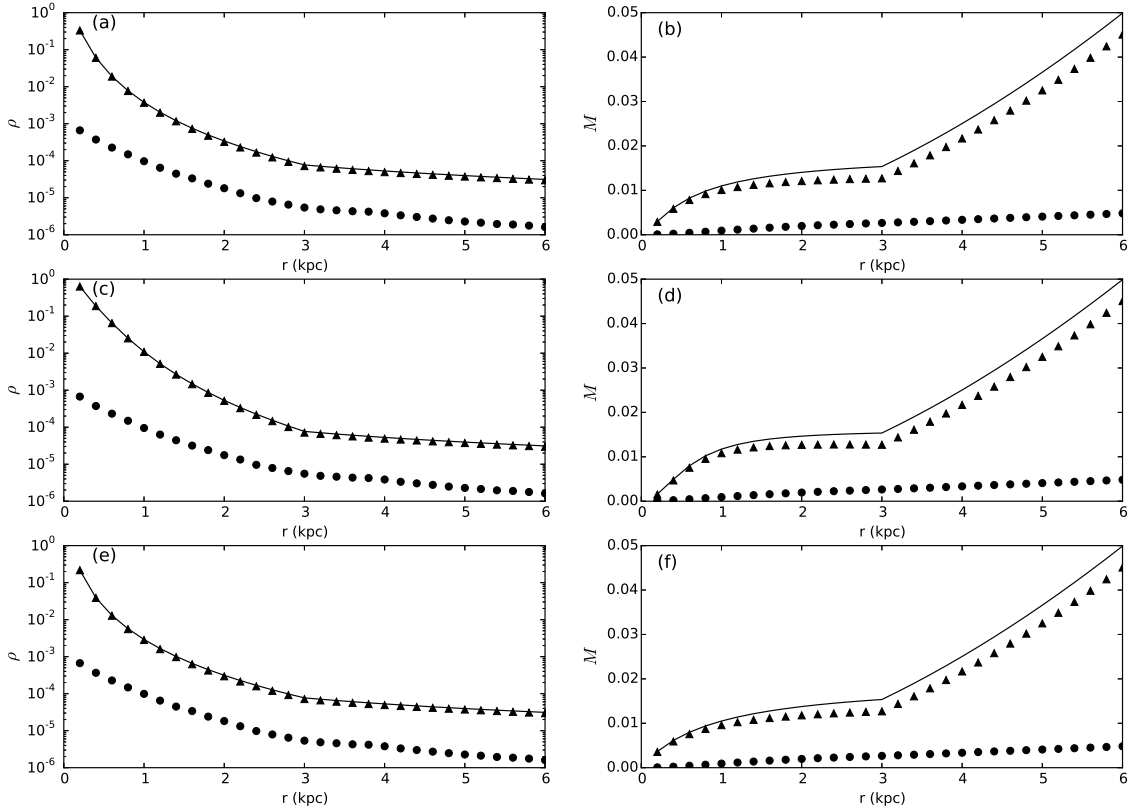}
\caption{The density and the cumulative mass as functions of radius $r$
for three best-fit models. 
Left panels are for 
the density and right panels are for the cumulative mass.
Panels (a)-(b) are for Model 1, Panels (c)-(d) are for Model 2, 
and Panels (e)-(f) are for Model 3.
In each panel, 
the triangles are for the dark matter, 
the circles are for the stellar part, 
and the solid curves are for the total. 
The density unit is $1.2\times 10^{12} M_{\odot}/({\rm {kpc}^3})$
and the mass unit is $1.2 \times 10^{12} M_{\odot}$. 
}
\label{fig:fig_9}
\end{figure}

\end{document}